# Particle collisions around static spherically symmetric black hole and rotating black hole in gravity's rainbow

Deng Lin-fang[1], Zhang He-Yao[2*] and Long Chao-Yun[3*]

1 College of Science, Changzhou Institute of Technology, Changzhou, 213032, China.

2 Wang Zheng School of Microelectronics, Changzhou University, Changzhou, 213164 China

3 College of Physics, Guizhou University, Guiyang 550025, China

**Abstract:** The gravity's rainbow is one well-studied modified theories of gravity based on a modified energy–momentum dispersion relation in the UV limit. Within this framework, the spacetime metric depends on the energy of the test particle. We extend the Banados-Silk-West effect to the static spherically symmetric black hole and rotating black hole in gravity's rainbow. Through systematic investigation that the effects of different rainbow functions on the center-of-mass energy of two test particles colliding outside the event horizon, we discussed the possibility of infinite center-of-mass energy $E_{cm}$ and the corresponding motion parameters. By employing rainbow functions $g_0 = g_1 = \left(1 - E/E_P\right)^{-1}$, infinite $E_{cm}$ can be achieved through collisions occurring outside the event horizon, rather than being confined solely to collisions near the event horizon.

1. Introduction

The Banados-Silk-West (BSW) effect, first proposed in 2009, showed that the rotating black hole can act as particle accelerators, producing arbitrarily high center-of-mass energy ($E_{cm}$) through a collision of two particles near horizon of an extremal Kerr black hole.[1] This phenomenon arises uniquely provided the critical angular momentum of one of the colliding particles, for geodesic particles. Subsequent work further elucidated the BSW mechanism and discussed practical limitations [2,3], and it was demonstrated that the BSW effect is one of the universal properties of rotating black holes [4,5]. It was later shown that the BSW mechanism is also possible for nonextremal black holes [6]. Due to that it can probe Planck-scale physics or particle physics that extends beyond the standard model, or study the energy extraction from a black hole [7-9], BSW effect aroused much attention which continues today. A series of subsequent studies extended the colliding particles to include charged [10,11], spinning[12,13], and accelerated particle[14], spinning magnetized particles[15], particles with magnetic dipole moment and electric charge[16,17], two different massive colliding particles[18], neutral particle with a charged particle[19]. In [20], the classification of near-fine-tuned particles (one of colliding particles has fine-tuned parameters) was given and all possible cases were shown. The BSW mechanism was also studied in various black holes, such as static spherically symmetric black hole[21], weakly magnetized black hole[19], Kerr-Taub-NUT black hole[22,23], rotating and accelerating black hole[24], rotating regular Hayward's black hole[25], Einstein-Maxwell-dilaton black hole[26], and recently, higher dimensional black hole[27], rotating charged black holes with Weyl corrections[28], rotating Simpson-Visser black hole[29], charged rotating black hole surrounded by perfect fluid dark matter[30], non-rotating Konoplya and Zhidenko black hole immersed in an external uniform magnetic field[31]. In recent years, with many modifications to general relativity were obtained, BSW effect in different gravity theories are also taking several attentions, such as the scalar-tensor-vector gravity[7,32,33], Kalb-Ramond gravity[34,35], Starobinsky-Bel-Robinson gravity[36], Einstein-Maxwell-scalar gravity[37].

While these particle collisions are all limited to regions near the event horizon, our study has revealed that within the framework of gravity's rainbow theory, infinitely $E_{cm}$ can still be achieved even when collisions do not occur close to the horizon. In this paper, we discuss $E_{cm}$ of two test particles colliding in spacetime modified by rainbow functions. in Sec. 2, we introduce the gravity's rainbow theory and rainbow function. In Sec. 3 and Sec. 4, we study collisions of two particles outside the event horizon of a static spherically symmetric black hole and rotating black holes in gravity's rainbow respectively.

2. the gravity's rainbow and rainbow function

As another modification of gravity, the gravity's rainbow is the extension of the doubly special relativity theory to curved spacetime.[38,39] In this formalism, the geometry of spacetime correlates with the probing particle's energy, so the modification of the metric by certain phenomenologically motivated rainbow functions causes different speeds of light with different wavelengths, called "Rainbow effect".[40] The gravity's rainbow has gathered a lot of attention, and it has been an inviting research field over the years. Recently, many different physical problems in the context of gravity's rainbow theory were studied such as the accelerating AdS black hole solutions[41], three-dimensional AdS gravitational vacuum stars[42], the dark energy star[43], the $f(R)$ models of inflation[44,45], the rainbow gravity effect on Klein-Gordon (KG) oscillators[46], thermodynamics of Schwarzschild black hole surrounded by quintessence[47], wormholes[48,49], etc. The modified metric can be evaluated using[38],

$$g(E) = \eta^{ab} e_a(E) \otimes e_b(E), \tag{1}$$

where, $e_0(E) = \frac{1}{g_0(E/E_P)} \tilde{e}_0$, $e_i(E) = \frac{1}{g_1(E/E_P)} \tilde{e}_i$. Here, $g_0(E/E_P)$ are $g_1(E/E_P)$ the Rainbow functions, $\tilde{e}_0$ and $\tilde{e}_i$ is the energy independent frame fields, $E$ and $E_P$ represent the energy of the system and Planck energy scale respectively. For relativistic particles and anti-particles, $0 \leq E/E_P \leq 1$ [46]. It means that the modified metric can get simply by making the change $dt \to dt/g_0(E/E_P)$ and all spatial coordinates $dx^i \to dx^i/g_1(E/E_P)$.

3. Collision around the static spherically symmetric black hole in gravity's rainbow

We discuss the collision of two particles around a static spherically symmetric black hole in gravity's rainbow. For a static and spherically symmetric spacetime, the modified metric is given by,

$$ds_i^2 = \frac{\tilde{g}_{tt}}{g_0^2(E_i/E_P)} dt^2 + \frac{\tilde{g}_{rr}}{g_1^2(E_i/E_P)} dr^2 + \frac{\tilde{g}_{\theta\theta}}{g_1^2(E_i/E_P)} d\theta^2 + \frac{\tilde{g}_{\varphi\varphi}}{g_1^2(E_i/E_P)} d\varphi^2, \tag{2}$$

where the indices refer to the *i*-th particle ( $i = 1, 2$ ). We will restrict our discussion to the equatorial plane ( $\theta = \frac{\pi}{2}$ ),

$$ds_i^2 = \frac{\tilde{g}_{tt}(r)}{g_0^2(E_i/E_P)}dt^2 + \frac{\tilde{g}_{rr}(r)}{g_1^2(E_i/E_P)}dr^2 + \frac{\tilde{g}_{\varphi\varphi}(r)}{g_1^2(E_i/E_P)}d\varphi^2 \qquad (3)$$
$$= g_{tt}(r,E_i)dt^2 + g_{rr}(r,E_i)dr^2 + g_{\varphi\varphi}(r,E_i)d\varphi^2$$

Next, we calculate the motion equation of particles, which will be necessary to determine the $E_{cm}$ of the two colliding particles. The motion of the particle is determined by the Lagrangian,

$$L_i = \frac{1}{2}g_{\mu\nu}(r,E_i)\dot{x}_i^\mu \dot{x}_i^\nu = \frac{1}{2}g_{tt}(r,E_i)\dot{t}_i^2 + \frac{1}{2}g_{rr}(r,E_i)\dot{r}_i^2 + \frac{1}{2}g_{\varphi\varphi}(r,E_i)\dot{\varphi}_i^2 . \qquad (4)$$

We can write the geodesic equations for a particle in terms of two conserved quantities, the energy *E* and the angular momentum *L*. The conserved quantities at the equatorial plane are defined by the following equations:

$$-E_i = \frac{\partial L_i}{\partial \dot{t}_i} = g_{tt}(r,E_i)\dot{t}_i$$
$$p_{\varphi i} = \frac{\partial L_i}{\partial \dot{\varphi}_i} = g_{\varphi\varphi}(r,E_i)\dot{\varphi}_i = \ell_i \qquad (5)$$

From the normalization condition

$$g_{\mu\nu}(r,E_i)\dot{x}_i^\mu \dot{x}_i^\nu = g_{tt}(r,E_i)\dot{t}_i^2 + g_{rr}(r,E_i)\dot{r}_i^2 + g_{\varphi\varphi}(r,E_i)\dot{\varphi}_i^2 = -1 . \qquad (6)$$

Then, the radial trajectory of the particle can be obtained by following equation,

$$\dot{r}_i^2 = -\frac{E_i^2}{g_{tt}(r,E_i)g_{rr}(r,E_i)} - \frac{1}{g_{rr}(r,E_i)}\left(\frac{\ell_i^2}{g_{\varphi\varphi}(r,E_i)} + 1\right) \equiv R_i(r,E_i) . \qquad (7)$$

The energy in the center of mass frame can easily obtain,

$$E_{cm}^2 = -P_\tau^a P_{\tau a} = -(m_1\dot{x}_1^a + m_2\dot{x}_2^a)(m_1\dot{x}_{1a} + m_2\dot{x}_{2a}) = (m_1^2 + m_2^2) - m_1m_2(\dot{x}_1^a\dot{x}_{2a} + \dot{x}_2^a\dot{x}_{1a}) \qquad (8)$$
$$= (m_1^2 + m_2^2) - m_1m_2\left[\dot{x}_1^a g_{ab}(r,E_2)\dot{x}_2^b + \dot{x}_2^a g_{ab}(r,E_1)\dot{x}_1^b\right]$$

We can get,

$$\frac{E_{cm}^2}{m_1m_2} = \frac{m_1^2 + m_2^2}{m_1m_2} - \left[g_{ab}(r,E_1) + g_{ab}(r,E_2)\right]\dot{x}_1^a \dot{x}_2^b . \qquad (9)$$

If $E_1 = E_2$, then $g_{ab}(r,E_1) = g_{ab}(r,E_2)$. Specifically, when the modification of the rainbow function is not considered $g_0 = g_1 = 1$, and $E_1 = E_2, m_1 = m_2 = m$, then the Eq.(9) becomes

$$\frac{E_{cm}^2}{2m^2} = 1 - 2\tilde{g}_{ab}(r)\dot{x}_1^a \dot{x}_2^b. \tag{10}$$

If collide at $r = r_p > r_H$, in the equatorial plane with the radial velocity equal to zero, $R_i(r) = 0$, we can get the $E_{cm}$ of two particles $m_1 = m_2 = m$, and it takes the following form,

$$\frac{E_{cm}^2}{2m^2} = 1 - \frac{1}{2}\frac{[g_{tt}(r_P, E_1) + g_{tt}(r_P, E_2)]}{g_{tt}(r_P, E_1) g_{tt}(r_P, E_2)} E_1 E_2 - \frac{1}{2}\frac{[g_{\varphi\varphi}(r_P, E_1) + g_{\varphi\varphi}(r_P, E_2)]}{g_{\varphi\varphi}(r_P, E_1) g_{\varphi\varphi}(r_P, E_2)} \ell_1 \ell_2 \tag{11}$$

Considering the modified metric of a static and spherically symmetric spacetime in the rainbow's gravity,

$$g_{tt}(r, E_i) = -\frac{f(r)}{g_0^2(E_i/E_P)}, \quad g_{rr}(r, E_i) = \frac{1}{f(r)g_1^2(E_i/E_P)}, \quad g_{\varphi\varphi}(r, E_i) = \frac{r^2}{g_1^2(E_i/E_P)}, \tag{12}$$

the Eq.(11) become,

$$\frac{E_{cm}^2}{2m^2} = 1 + \frac{1}{2}\frac{[g_0^2(E_1/E_P) + g_0^2(E_2/E_P)]}{f(r_P)} E_1 E_2 - \frac{1}{2}\frac{[g_1^2(E_1/E_P) + g_1^2(E_2/E_P)]}{r_P^2} \ell_1 \ell_2. \tag{13}$$

Consequently, we employ the rainbow functions which was considered in [50-53],

$$g_0(E/E_P) = g_1(E/E_P) = (1 - E/E_P)^{-1} \tag{14}$$

The choices of the rainbow function due to the theory with constant velocity of light and also solves the horizon problem. In this case, from Eq.(13), we can find that the center of mass energy diverges, $\frac{E_{cm}^2}{2m^2} \to \infty$, when the energy of at least one particle $E_i \to E_P$ and $\ell_1 \ell_2 < 0$. That is, the energy of the particles after collision approaches infinity, and the collision occurs at $r_p > r_H$. In the Appendix-A, we discuss the modifications of other rainbow functions and find no divergent center-of-mass energy except for collisions near the event horizon.

According to Eq. (7), when $R_i(r) = 0$, we can get,

$$\frac{E_i^2}{g_{tt}(r_p)} = -\left(\frac{\ell_i^2}{g_{\varphi\varphi}(r_p)} + 1\right). \tag{12}$$

The solution is $\ell^2 = \left[E^2 + (1 - E/E_P)^2\right]\frac{r_p^2}{f(r_p)}$.

## 4. Collision around the rotating black hole in gravity's rainbow

Next, we discuss particle collisions around rotating black holes. The modified line element for a rotating black hole is given by,

$$ds_i^2 = \frac{\tilde{g}_{tt}}{g_0^2(E_i/E_P)}dt^2 + \frac{2\tilde{g}_{t\varphi}}{g_0(E_i/E_P)g_1(E_i/E_P)}dtd\varphi + \frac{\tilde{g}_{rr}}{g_1^2(E_i/E_P)}dr^2 + \frac{\tilde{g}_{\theta\theta}}{g_1^2(E_i/E_P)}d\theta^2 + \frac{\tilde{g}_{\varphi\varphi}}{g_1^2(E_i/E_P)}d\varphi^2. \quad (16)$$

In the equatorial plane ($\theta = \frac{\pi}{2}$),

$$ds_i^2 = \frac{\tilde{g}_{tt}(r)}{g_0^2(E_i/E_P)}dt^2 + \frac{2\tilde{g}_{t\varphi}(r)}{g_0(E_i/E_P)g_1(E_i/E_P)}dtd\varphi + \frac{\tilde{g}_{rr}(r)}{g_1^2(E_i/E_P)}dr^2 + \frac{\tilde{g}_{\varphi\varphi}(r)}{g_1^2(E_i/E_P)}d\varphi^2$$
$$= g_{tt}(r,E_i)dt^2 + 2g_{t\varphi}(r,E_i)dtd\varphi + g_{rr}(r,E_i)dr^2 + g_{\varphi\varphi}(r,E_i)d\varphi^2 \quad (17)$$

Using the Lagrangian Eq.(4), the total angular momentum and total energy of the particle take the form,

$$E_i = \frac{\partial L_i}{\partial \dot{t}_i} = g_{tt}(r,E_i)\dot{t}_i + g_{t\varphi}(r,E_i)\dot{\varphi}_i,$$
$$\ell_i = \frac{\partial L_i}{\partial \dot{\varphi}_i} = g_{t\varphi}(r,E_i)\dot{t}_i + g_{\varphi\varphi}(r,E_i)\dot{\varphi}_i. \quad (18)$$

Then, due to $g_{\mu\nu}(r,E_i)\dot{x}_i^\mu \dot{x}_i^\nu = -1$, the equation of radial motion can be outlined by,

$$\dot{r}_i^2 = \frac{1}{g_{rr}(r,E_i)}\left[\frac{E_i^2 g_{\varphi\varphi}(r,E_i) + 2E_i\ell_i g_{t\varphi}(r,E_i) + \ell_i^2 g_{tt}(r,E_i)}{\left(g_{t\varphi}^2(r,E_i) - g_{tt}(r,E_i)g_{\varphi\varphi}(r,E_i)\right)} - 1\right] \equiv R_i(r). \quad (19)$$

We restrict the path at the turning point $R_i(r) = 0$, then $E_{cm}$ of the two particles with $m_1 = m_2 = m$ is given by,

$$\frac{E_{cm}^2}{2m^2} = 1 + \frac{1}{2}\left[\frac{g_{\varphi\varphi}(E_1)}{g_{t\varphi}^2(E_1) - g_{tt}(E_1)g_{\varphi\varphi}(E_1)} + \frac{g_{\varphi\varphi}(E_2)}{g_{t\varphi}^2(E_2) - g_{tt}(E_2)g_{\varphi\varphi}(E_2)}\right]E_1 E_2$$
$$+ \frac{1}{2}\left[\frac{g_{tt}(E_1)}{g_{t\varphi}^2(E_1) - g_{tt}(E_1)g_{\varphi\varphi}(E_1)} + \frac{g_{tt}(E_2)}{g_{t\varphi}^2(E_2) - g_{tt}(E_2)g_{\varphi\varphi}(E_2)}\right]\ell_1 \ell_2 \quad (20)$$
$$+ \frac{1}{2}\left[\frac{g_{t\varphi}(E_1)}{g_{t\varphi}^2(E_1) - g_{tt}(E_1)g_{\varphi\varphi}(E_1)} + \frac{g_{t\varphi}(E_2)}{g_{t\varphi}^2(E_2) - g_{tt}(E_2)g_{\varphi\varphi}(E_2)}\right](E_1\ell_2 + E_2\ell_1)$$

when $E_1 = E_2 = E$, $\ell_1 = -\ell_2 = \ell$, $E_{cm}$ can be written in the following form,

$$\frac{E_{cm}^2}{2m^2} = 1 + \frac{g_{\varphi\varphi}(E)}{g_{t\varphi}^2(E) - g_{tt}(E)g_{\varphi\varphi}(E)}E^2 - \frac{g_{tt}(E)}{g_{t\varphi}^2(E) - g_{tt}(E)g_{\varphi\varphi}(E)}\ell^2. \quad (21)$$

We discuss the Kerr metric for a spinning, spherical mass, and this metric is as follows,

$$\tilde{g}_{tt} = \frac{2M}{r} - 1, \tilde{g}_{t\varphi} = -\frac{2aM}{r}, \tilde{g}_{\varphi\varphi} = \left(r^2 + a^2 + \frac{2a^2 M}{r}\right). \tag{22}$$

For the radius of the Kerr black hole's horizon $r_{\pm} = M \pm \sqrt{M^2 - a^2}$, we can get $r_+ < 2M$. Let us choose the point of collision with $r_p \geq r_+ = 2M$. Subsequently, substituting the rainbow function $g_0(E/E_P) = g_1(E/E_P) = (1 - E/E_P)^{-1}$ into Eq. (20), we get

$$\frac{E_{cm}^2}{2m^2} = 1 + \frac{1}{2}\left[\frac{1}{(1-E_1/E_P)^2} + \frac{1}{(1-E_2/E_P)^2}\right]\frac{\left[\tilde{g}_{\varphi\varphi}(r_P)E_1 E_2 + \tilde{g}_{tt}(r_P)\ell_1\ell_2 + \tilde{g}_{t\varphi}(r_P)(E_1\ell_2 + E_2\ell_1)\right]}{\tilde{g}_{t\varphi}^2(r_P) - \tilde{g}_{tt}(r_P)\tilde{g}_{\varphi\varphi}(r_P)}. \tag{23}$$

As Eq. (23) show, if the energy of any one of these particles $E_i \to E_p$, a divergent center-of-mass energy can be obtained. Solving the condition $R_i(r) = 0$, the angular momentum becomes,

$$\ell^2 = \left[-2E\tilde{g}_{t\varphi} + \sqrt{4E^2\tilde{g}_{t\varphi}^2 + 4(\tilde{g}_{t\varphi}^2 - \tilde{g}_{tt}\tilde{g}_{\varphi\varphi})(1-E/E_P)^2 - 4E^2\tilde{g}_{\varphi\varphi}}\right]/2\tilde{g}_{tt}\bigg|_{r=r_P}. \tag{25}$$

4. Conclusion

We investigated the center of mass energy of particle collisions under rainbow function modified metrics near both generic static black holes and rotating black holes. Our results reveal that for the rainbow function $g_0 = g_1 = (1 - E/E_P)^{-1}$, when at least one particle possesses a critical energy $E_i \to E_P$, the $E_{cm}$ diverges to infinity due to the inherent divergence of rainbow function. Moreover, this occurs for collisions at positions where $r_p > r_H$, requiring no close to the event horizon. In contrast, the other two rainbow functions consistently ensure finite $E_{cm}$.

Appendix-A

Here, let us consider the other rainbow functions. For the rainbow functions proposed by Amelino-Camelia et al. in [54,55],

$$g_0 = 1, \, g_1 = \sqrt{1 - \eta(E/E_P)^n}, \quad n = 1, 2. \tag{A-1}$$

These rainbow functions are compatible with results from with loop quantum gravity and non-commutative geometry[56]. Then $E_{cm}$ of particle around the static spherically symmetric black hole is determined

$$\frac{E_{cm}^2}{2m^2} = 1 + \frac{1}{f(r)} E_1 E_2 - \frac{1}{2} \frac{\left[2 - \eta(E_1/E_P)^n - \eta(E_2/E_P)^n\right]}{r^2} \ell_1 \ell_2 \qquad (A-2)$$

For a rotating black hole, Eq. (23) becomes,

$$\frac{E_{cm}^2}{2m^2} = 1 + \frac{\tilde{g}_{\varphi\varphi}(r)}{\tilde{g}_{t\varphi}^2(r) - \tilde{g}_{tt}(r)\tilde{g}_{\varphi\varphi}(r)} E_1 E_2 + \frac{1}{2} \frac{\tilde{g}_{tt}(r)}{\tilde{g}_{t\varphi}^2(r) - \tilde{g}_{tt}(r)\tilde{g}_{\varphi\varphi}(r)} \left[2 - \eta(E_1/E_P)^n - \eta(E_2/E_P)^n\right] \ell_1 \ell_2$$
$$+ \frac{1}{2} \frac{\tilde{g}_{t\varphi}(r)}{\tilde{g}_{t\varphi}^2(r) - \tilde{g}_{tt}(r)\tilde{g}_{\varphi\varphi}(r)} \left[\sqrt{1 - \eta(E_1/E_P)^n} - \sqrt{1 - \eta(E_2/E_P)^n}\right](E_1\ell_2 + E_2\ell_1) \qquad (A-3)$$

The third choice for the rainbow function obtained from the spectra of gamma-ray bursts [57], given as,

$$g_0 = 1 - e^{-\eta(E/E_P)}, \quad g_1 = 1. \qquad (A-4)$$

It has only partially complied with the intended rainbow gravity effect [58]. The center of mass energy for the static spherically symmetric black hole is given by following form,

$$\frac{E_{cm}^2}{2m^2} = 1 + \frac{1}{2} \frac{\left[\left(1 - e^{-\eta(E_1/E_P)}\right)^2 + \left(1 - e^{-\eta(E_2/E_P)}\right)^2\right]}{f(r)} E_1 E_2 - \frac{1}{r^2} \ell_1 \ell_2. \qquad (A-5)$$

For particle around a rotating black hole, $E_{cm}$ as following,

$$\frac{E_{cm}^2}{2m^2} = 1 + \frac{1}{2} \frac{\tilde{g}_{\varphi\varphi}(r)}{\tilde{g}_{t\varphi}^2(r) - \tilde{g}_{tt}(r)\tilde{g}_{\varphi\varphi}(r)} \left[\left(1 - e^{-\eta(E_1/E_P)}\right)^2 + \left(1 - e^{-\eta(E_2/E_P)}\right)^2\right] E_1 E_2 + \frac{\tilde{g}_{tt}(r)}{\tilde{g}_{t\varphi}^2(r) - \tilde{g}_{tt}(r)\tilde{g}_{\varphi\varphi}(r)} \ell_1 \ell_2$$
$$+ \frac{1}{2} \frac{\tilde{g}_{t\varphi}(r)}{\tilde{g}_{t\varphi}^2(r) - \tilde{g}_{tt}(r)\tilde{g}_{\varphi\varphi}(r)} \left[2 - e^{-\eta(E_1/E_P)} - e^{-\eta(E_2/E_P)}\right](E_1\ell_2 + E_2\ell_1) \qquad (A-6)$$

In consideration of $0 \leq 1 - \eta(E/E_P)^n \leq 1$, and $0 \leq 1 - e^{-\eta(E/E_P)} \leq 1$, the center of mass energy shown by Eq.(A-2), Eq.(A-3), Eq.(A-5) and Eq.(A-6) are always finite.

**Acknowledgements** This research is also supported by Grant No. E4-6111-25-034 of the Changzhou Institute of Technology. D.L. and Z.H. would like to express their gratitude for the birth of their daughter Zhang Xun-Yu, because that, with her arrival, they saw a rainbow emerging in the post-typhoon sky.

**Data Availability Statement** My manuscript has no associated data. [Author's comment: This is a theoretical study without experimental or observational data.]

**Code Availability Statement** My manuscript has no associated code/software. [Author's comment: This paper is a theoretical study, and no associated code/software is involved.]